\documentstyle[a4,12pt,epsfig]{article}

\title
 {\bf Study of in-medium $\omega$ meson properties
in Ap, pA and AA collisions}
\author{S.Belogurov, M.Chumakov, S.Kiselev,
Yu.Kiselev\footnote{e-mail address: yurikis@itep.ru}, V.Sheinkman}

\begin{document}

\maketitle

\begin{center}
{\small {\it Institute for Theoretical and Experimental Physics

Moscow, Russia\\
}}
\end{center}
\begin{abstract}
We propose to investigate the in-medium properties of vector
$\omega$ mesons at the normal nuclear density in Ap(pA) collisions and
at higher density in AA collisions at the ITEP accelerator facility
TWAC. Using of the inverse Ap kinematics will permit us to study
the $\omega$ meson production in a wide momentum interval included the
not yet explored range of small meson momenta relative to the
projectile nuclei where the mass modification effect in nuclear
matter is expected to be the strongest. Momentum dependence of the
in-medium $\omega$ meson width will be studied in the traditional pA
kinematics. We intend to use the electromagnetic calorimeter for 
reconstruction of the $\omega$ meson invariant mass by detecting 
photons from the $\omega \rightarrow \pi^{0}\gamma \rightarrow
3\gamma$ decay. The model calculations and simulations with RQMD
generator show feasibility of the proposed experiment. Available
now intensity of the ion beams provides a possibility to collect
large statistics and make decisive conclusion about the $\omega$
meson properties at density of normal nuclei. At the second
stage of the investigation the $\omega$ meson properties will be
studied in AA collisions at higher density. Interpretation of
these measurements will be based on the results obtained in Ap(pA)
interactions. Further investigation of the in-medium properties of
light unflavored and charmed mesons can be performed at ITEP and at
GSI(FAIR) where higher ion energies will be accessible in near
future.
\end{abstract}

\newpage
\tableofcontents

\newpage

\section{Physics motivaion}

Modification of the hadron properties in baryon environment is
one of the important topics of contemporary strong interaction
physics. This phenomenon has been predicted within various
theoretical approaches such as QCD sum rules~\cite{QCDsumRules},
chiral dynamics~\cite{ChiralDynamics}, relativistic
mean-field~\cite{RelMeanField} and quark-meson coupling
model~\cite{QMCM}. A recent review can be found in the
Ref.~\cite{Review}. A hadron can change its properties such as mass
and width once it is embedded into a baryon matter. This change is
connected to the many body interactions of a hadron with surrounding
nucleons. Whether a hadron is - in addition - also affected by QCD
condensates and their in-medium change ~\cite{QCDsumRules},
~\cite{ChiralDynamics}, ~\cite{Debate} is still a matter of debate.
Nevertheless, the great interest in study of in-medium hadron
properties is caused by the expectation to find the evidences of the
chiral symmetry restoration. Investigation of the vector mesons
is of special interest in this context. Theoretically, the
possibility of the decrease in the mass of light vector mesons in
matter was first pointed  in \cite{Bernard} and later in~\cite{ChiralDynamics}.
According to~\cite{ChiralDynamics}  the masses of the vector mesons scale with quark
condensate, i.e. drop with rising of baryonic density. This effect can be
a  precursor phenomenon of the
transition of strongly interacting matter to the chirally symmetric
phase. First experimental signal of this phenomenon was recently
observed in ~\cite{chsymrest}. Nambu and Jona-Lasino proposed the
spontaneous breaking of the chiral symmetry as the fundamental
mechanism for the creation of a mass of hadrons~\cite{NJL}.
Recently, the in-medium change of the $\omega$ mesons spectral
function was proposed as a probe of higher order QCD four-quark
condensate~\cite{FourQuarkCond}.

An evidence for a decrease of the $\rho$ meson mass in heavy-ion
collisions was obtained by the CERES collaboration at CERN
~\cite{CERES} and later by the STAR collaboration at
RHIC~\cite{STAR}. Since heavy-ion interaction is very complicated
process in which the temperature and baryon density varies
dramatically with time due to the formation and expansion of the
"fireball", the interpretation of experimental data on
nucleus-nucleus collisions is far from being simple. The above
results have been found an explanation in terms of shifting a $\rho$
meson spectral function to a lower mass, as expected from the
theory. However, even the calculations that just used the free
radiation rates with their - often quite large - experimental
uncertainties are compatible with the observation.

Therefore, it is useful to explore the reactions with elementary
probes ($\gamma$, $\pi$, p) since sizeable - about 20\% - medium
effects were predicted already at the density of ordinary
nuclei~\cite{ChiralDynamics},~\cite{20persent1},~\cite{20persent2}.
The advantage of the investigations of the reactions on nuclei is
related to the fact that they proceed in the nearly cold static
nuclear matter and thus the colliding system is much better under
control. Indeed, the first signals for lowering of the ${\omega}$
meson mass at normal nuclear matter density were recently observed
in the $\gamma A$ ~\cite{ELSA} and $pA$ ~\cite{KEK} reactions.
However, the critical analysis \cite{Kaskulov} shows that data of
the experiment \cite{ELSA} are compatible with normal $\omega$ mass
and an enlarged  width. In contrast to the conclusion ~\cite{KEK}
the preliminary results of the CLAS collaboration (JLAB) on the
photoproduction of $\rho$ and $\omega$ mesons ~\cite{JLAB} also
evidence for no shift in the mass. Now there are only first
estimates of the $\omega$ meson width in matter ~\cite{ELSA}, ~\cite
{Kotulla}. Thus, the available now experimental information does not
allow to draw the final conclusion about the change of the $\omega$
meson properties even in nuclear matter of normal density. It should
be stressed that the indications for decreasing of the $\omega$
meson mass in both experiments ~\cite{ELSA},~\cite{KEK} have been
found for the mesons with low momenta relative to the surrounding
nuclear matter. Therefore, next generation of experiments need to
addresses the issue of momentum dependence of medium effects. We
suggest to explore the momentum dependence of the in-medium mass and
width of the $\omega$ meson using the ion and proton beams of the
ITEP accelerator facility TWAC~\cite{ITEP-TWAC}.

The investigation of in-medium meson modification addresses the
fundamental problems of strong interaction physics and is one of the
hot current topics nowadays. The experiments with photon, pion,
proton and ion projectiles are planned in wide collision energy
range from a few GeV (GSI, JLAB, JINR, COSY, SPring-8, ITEP) till
TeV (RHIC, LHC).

\section{Goal of the experiment}
The goal of the proposed experiment is the investigation of the
vector $\omega$ meson properties at normal nuclear density
$\rho_{0}=0.17 fm^{-3}$ in nucleus-proton (proton-nucleus)
collisions and at higher density in nucleus-nucleus collisions. The
experiment aims at the study of the mesons with low momentum
relative to the baryonic environment where the in-medium mass
modification is expected to be most strong as well as at the study
of high momentum range which is sensitive to the in-medium $\omega$
meson width.

\section{Theoretical predictions}
All information about the intrinsic properties of a meson is encoded
in its spectral function S(M) which can be written in
non-relativistic Breit-Wigner form. In free space:
\begin{equation}
\label{eq:BreitWigner}
      S(M) = (\Gamma_{0}/2)^{2}/[(M-M_{0})^{2} + (\Gamma_{0}/2)^{2}],
\end{equation}
where $\Gamma_{0}$ and
$M_{0}$ stand for a meson width and pole mass, correspondingly.

Due to the interaction with surrounding nuclear medium the meson
acquires a selfenergy $\Sigma$ which is related to the nuclear
optical potential U as~\cite{Cabrera}:
\begin{equation}
\label{eq:OptPot}
      \Sigma/2E = U = ReU + {\it i}ImU,
\end{equation}
where E is the total meson energy.

The meson spectral function in nuclear medium is read:
\begin{equation}
\label{eq:SpectrFunction}
    S(M) = [(\Gamma_{0}/2)+(\Gamma^{*}/2)]^{2}/[M-(M_{0}+ M^{*})]^{2} + [(\Gamma_{0}/2)+(\Gamma^{*}/2)]^{2}.
\end{equation}
Two extra terms, $M^{*}$ and $\Gamma^{*}/2$, which
describe the shift of the meson pole mass and
the increase of its width in matter, are
related to the nuclear optical potential $U$ as follows~\cite{Cabrera}:
\begin{equation}
\label{eq:Param}
             M^{*} = ReU;             \Gamma^{*}/2 = - ImU;
\end{equation}
The pole mass and width of the $\omega$ meson in free space (vacuum)
are M = 782 MeV and 8.4 MeV, correspondingly. Most theoretical
investigations predict the dropping of the in-medium $\omega$ meson
mass by 20-140 MeV~\cite{TheorMass} at normal nuclear density.
However, there have also been suggestions for a rising
mass~\cite{RisingMass} or even a structure with several
peaks~\cite{SplitMass}. At the same time there seems to be a general
agreement that in-medium $\omega$ width is within the range from 20
MeV to 60 MeV~\cite{TheorWidth} at the density $\rho$ = $\rho_{0}$.
Thus, it is expected that the $\omega$ meson in matter survives as a
quasiparticle and can be observed as a structure in the $\omega$
mass spectrum. In principle, both dilepton and $\pi^{0}\gamma$
invariant mass spectra can be used for the study of modification
effects. The advantage of the dilepton decay channel is related to
the fact that leptons are almost undistorted by the final state
interactions. However, the $\omega$ signal in the dilepton mode is
rather weak ($BR(\omega \rightarrow e^{+}e^{-}) \approx 7.1 \times
10^{-5}$) and is always accompanied by a comparatively large
background from $\rho^{0} \rightarrow e^{+}e^{-}$ decays. The
$\omega \rightarrow \pi^{0}\gamma$ decay has a branching ratio $8.9
\times 10^{-2}$ what is 3 orders of magnitude higher. Furthermore,
the competing $\rho \rightarrow \pi^{0}\gamma$ channel has a
branching ratio which is a factor $10^{2}$ smaller. By these reasons
the $\omega \rightarrow \pi^{0}\gamma$ decay mode can be considered
as an exclusive probe to study the $\omega$ meson properties in
matter. The disadvantage of this channel is a possible rescattering
of the $\pi^{0}$ within the nuclear medium which would distort the
deduced $\omega$ invariant mass distribution.  However, as it was
shown in Refs.~\cite{EPJA20},~\cite{BUU} the above distortion effect
can be significantly decreased by applying an appropriate cut on the
pion kinetic energy.

\section{Inverse and direct kinematics}
The $\omega$ meson invariant mass spectrum has two components which
correspond to the decay 'inside' and 'outside' the nucleus. Only
vector mesons decaying 'inside' nuclei can be used for an
identification of the in-medium $\omega$ mass. This imposes the
kinematical condition that the decay length of the vector meson
should be less than nucleus size. It implies that the $\omega$ meson
should be produced with small momentum (velocity) relative to the
nuclear matter rest frame. The study of low momentum $\omega$ mesons
production in the inverse Ap kinematics~\cite{InverseKinem} has
several important advantages over the study in the direct pA
kinematics. First, as it follows from the Lorentz transformation,
slow particles in a projectile nucleus system appear to be fast in
the laboratory (in the target proton rest frame) and
 become convenient for the detection. At beam energy of 4 AGeV all the $\omega$'s produced in
full solid angle with momenta less than 0.3 GeV/c relative to the
projectile nucleus rest frame will be concentrated in the laboratory
inside narrow cone of less than $\pm5^{0}$ and the momentum range
from 2.8 till 5.9 GeV/c. The produced mesons which are almost at
rest inside the incident nucleus ("comovers") have the laboratory
momenta around of 4 GeV/c. Due to the decrease of the production
cross section with laboratory $\omega$ meson momentum the main
contribution to the $\omega$ yield comes from the momentum interval
of 2.8 - 4.0 GeV/c. These events will be observed in small phase
space $dPdcos\theta$ in the laboratory resulting in significant
increase in forward production cross section as compared to one in
pA reactions. That can be easily understood because experimentally
observed {\sl non-invariant double differential cross sections}
measured in the direct (pA) and inverse (Ap) kinematics are related
as:
\begin{equation}
\label{eq:DiffCrossSec}
     ^{Ap}(d^{2}\sigma/dPdcos\theta) = ^{Ap}(P^{2}/E)^{pA}(E/P^{2})^{pA}(d^{2}\sigma/dPdcos\theta).
\end{equation}
One can see that the factor $^{pA}(E/P^{2})$ grows strongly with
lowering of the $\omega$ meson momentum while the factor
$^{Ap}(P^{2}/E)$ changes rather smoothly.

The photons from the decay $\omega \rightarrow
\pi^{0}\gamma \rightarrow 3\gamma$ are distributed inside more wide
cone as compared to parent mesons, however the coverage of the
angular interval $5^{0}-25^{0}$ -  which corresponds to the solid
angle of less than 9\% of 4$\pi$  - permits to collect significant
part of the useful events.

Second, the mean free pass of the proton in nuclear matter is as
small as 2 fm and therefore the $\omega$ mesons are predominantly
created inside the front layers of a projectile nucleus. Since the
forward produced $\omega$'s in the momentum range 2.8-4.0 GeV/c have
the laboratory velocities which are less than ones of the
surrounding nucleons, the produced mesons move in the direction
opposite to the ion beam direction and then decay in more dense
inner layers of a nucleus. That is of great importance because the
strength of the medium effects increases with nuclear density.

Third advantage of the inverse kinematics is an increase in the
energies of the detected photons because they are emitted by
relativistic $\omega$ and $\pi^{0}$. For example, the
$\pi^{0}\gamma$ decay in transverse direction of the $\omega$
carrying the momentum of 4 GeV/c results in emission of the photon
of energy 2 GeV and $\pi^{0}$ of energy 2.1 GeV followed by the pion
decay to two photons of 1 GeV energy. The above energies exceed the
photon energies from the $\omega \rightarrow \pi^{0}\gamma$  decay
at rest (0.38 GeV for the $\gamma$ from $\omega$ and 0.19 GeV for
the $\gamma$'s from $\pi^{0}$) by a factor of about 5. That results
in more precise measurement of the photon energy leading to more
narrow width of the signal in the invariant mass spectrum and hence
improved signal to background ratio. At last, only moderate momentum
resolution in the laboratory is required for the rather precise
determination of the $\omega$ momentum relative to the projectile
nucleus because the momentum range of interest 2.8-4 GeV/c in the
laboratory corresponds to the interval 0-0.3 GeV/c in the nucleus
frame of reference.

In contrast with in-medium $\omega$ meson mass the value of its
width is expected to be deduced from the analysis of the production
of fast mesons relative to the baryonic matter. It is well known
that high momentum mesons can be abundantly produced in the pA
interactions. Thus, the combination of the Ap and pA measurements
provides the possibility to study both $\omega$ meson mass and width
in nuclear matter.

\section{Experimental arrangement}

\subsection{Extracted ion and proton beams}
\begin{figure}
\begin{center}
\includegraphics[width=12cm]{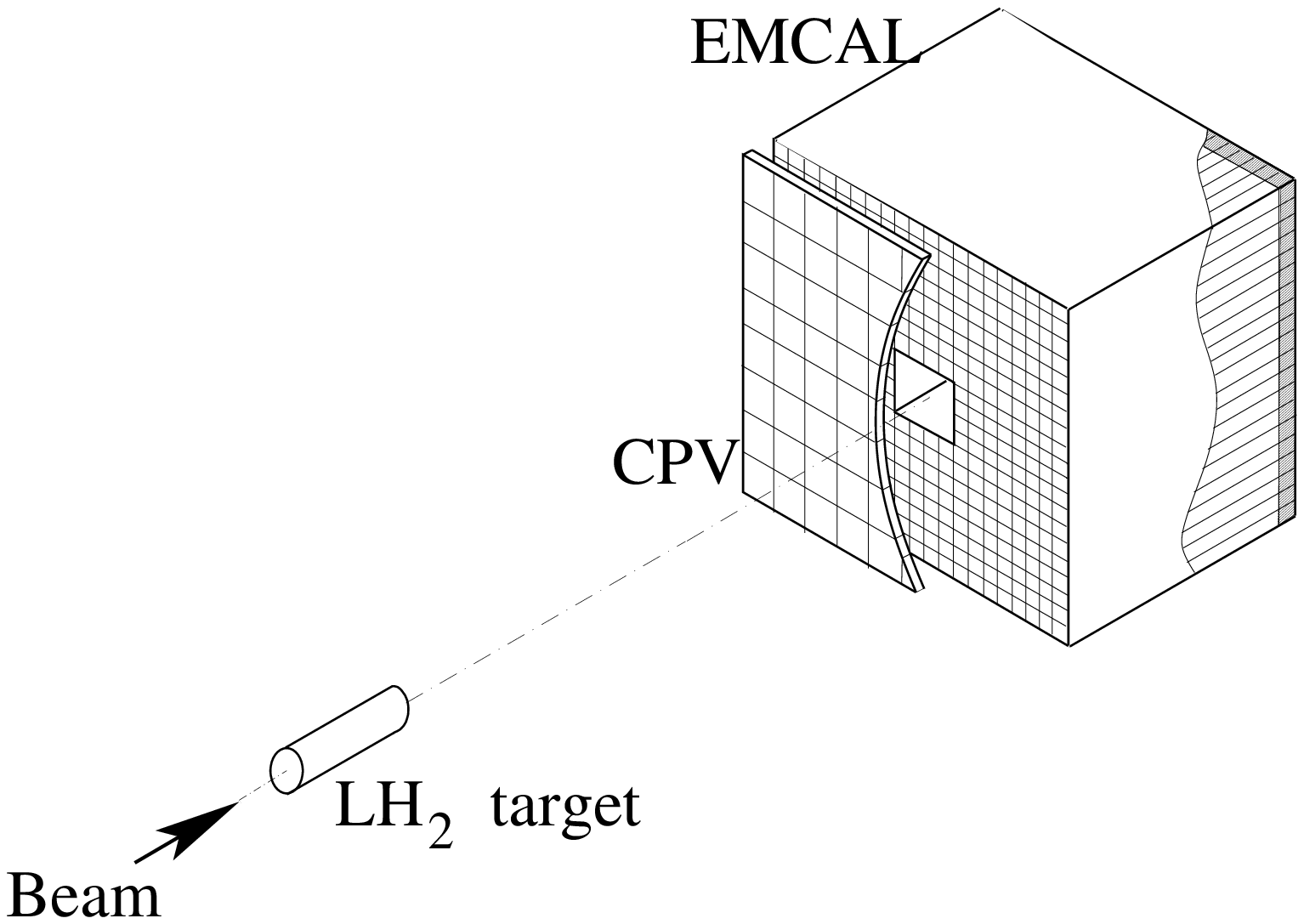}
\caption{Sketch of the experimental set-up}
\label{Fig Omsetup}
\end{center}
\end{figure}

    We intend to carry out the proposed measurements using the ion and proton beams
extracted in the inner hall of the accelerator. Expected extraction
efficiency is of 50\%. Two dipole and two pairs of quadrupole
magnets serve for the deflection and focusing the beams onto the
target. The sketch of the experimental set-up is shown in
Fig.~\ref{Fig Omsetup}. The ions (or protons) which do not interact
in the target pass through the central hole of the electromagnetic
calorimeter (EMCAL) and then directed to the downstream beam-dump
located in the inner hall or thick concrete wall of the accelerator.
That prevents the environment from the radiation pollution.

\subsection{Projectiles and targets}
    The projectile Ta, Cu, Al and C ions will be used for the investigations of the
in-medium $\omega$ meson mass while the projectile protons will be
used for the exploration of the $\omega$ width in the nuclear
matter. The proton runs will permit us to study the EMCAL
performance and calibrate the invariant mass scale by measurements
of the reactions $p+A\rightarrow\pi^{0}$+X, $p+A\rightarrow\eta$+X
and $p+A\rightarrow\omega$+X.

We plan to use the liquid hydrogen ($LH_{2}$) target of 2\%
interaction length (12 cm) for Ap measurements and simple foil
targets (Be, Al, Cu, Ag and Ta) for the pA and AA measurements.

\subsection{Photon detector}
The ring-like electromagnetic calorimeter with total area of 0.64
$m^{2}$ will be located at the distance of 1 $m$ downstream of the
target. We intend to use the EMCAL based on the PbWO cells $20
\times 20 mm^{2}$ size with avalanche photodiode or photomultiplier
readout. The energy and spatial resolution of the cell are
$\sigma$/E=2\%$\sqrt{E}$+1\% and $\sigma_{x} = \sigma_{y}=6 mm$,
respectively~\cite{Novotny}. The total number of cells is 1400. In
front of a group of cells the 5 mm thick plastic scintillator with
photodiode readout will be mounted for the detection of charged
particles. Due to the moderate charge ejectile multiplicity (see
section~\ref{simulations}) the number of CPV (Charged Particle Veto)
counters is less than about 100. This array can be also used as a
multiplicity detector offering the possibility to apply the cuts on
the impact parameter of the collision.

\section{Study of the $\omega$ meson in nuclear matter}

\subsection{In-medium $\omega$ meson mass}

For the evaluation of the expected signal of in-medium $\omega$
meson mass and width modification the calculations of the $\omega$
meson production were performed in the framework of the folding
model. The model takes properly into account both incoherent direct
proton-nucleon and secondary pion-nucleon $\omega$ meson production
processes as well as internal nucleon momentum distribution (see for
example~\cite{FoldingModel}). The folding model describes the
production, propagation and decay of the $\omega$ meson inside a
nucleus taking into account its four-momentum and local nuclear
density. The calculations were performed for Ta, Cu, Al and C nuclei
at initial energy of 4 AGeV.

In our approach the $\omega$ meson mass shift was introduced
according to the local nuclear density $\rho(r)$:

\begin{equation}
\label{eq:Mmed}
       M^{*} = ReU = \delta M_{0}\rho/\rho_{0},
\end{equation}

where $M_{0}$ stands for the $\omega$ meson vacuum mass. The
negative value of $\delta= -0.12$ - in accordance with the
theoretical predictions and the experimental observations
~\cite{ELSA},~\cite{KEK} - means that the $\omega$ meson feels a
strong attraction inside nuclear matter which is of about 90 MeV at
nuclear saturation density $\rho_{0}$. In our calculations the
nuclear density distributions were taken in two-parameter Fermi
form.

We primarily focus at study of the $\omega$ mesons with low momentum
in the projectile nucleus rest frame by two reasons. First, the
strength of the 'inside' component of the $\omega$ decay - which
carries the information on the in-medium meson mass - obviously
increases with lowering of a meson momentum. Second, the in-medium
$\omega$ mass shift can depend on the meson velocity with respect to
the surrounding nuclear matter (see discussion in ~\cite{EPJA20}).
The most strong effect is predicted to be manifest itself in the low
momentum range. One can also expect that the low momentum $\omega$
mesons can be captured by the nucleus which leads to the formation
of the $\omega$ - nucleus bound state ~\cite{Theory},
~\cite{Review}.

\begin{figure}
\begin{center}
\includegraphics[width=12cm]{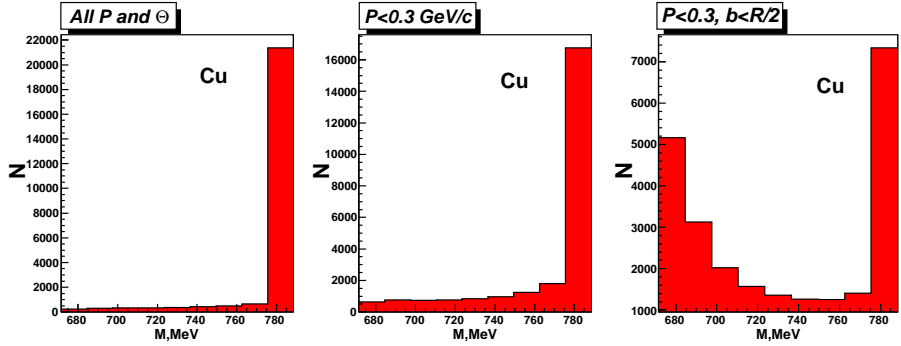}
\caption{$\omega$ meson mass spectra without and with the cuts on
the meson momentum and collision impact parameter}
\label{Fig_Om_Inv_Mass}
\end{center}
\end{figure}

The mass distributions of the $\omega$ mesons from Cu+p collisions
at 4 AGeV calculated within the frame of the folding model are
presented in Fig.~\ref{Fig_Om_Inv_Mass}.
In the left panel we show the mass spectrum of all produced
$\omega$'s. The right peak corresponds to the decays 'outside' the
nucleus and hence to the vacuum $\omega$\ meson mass while the left
part of the distribution corresponds to the decays 'inside' the
nucleus and contains the events with reduced meson
masses\footnote{We refer a decay to the 'inside' component provided
the local density $\rho/\rho_{0}>$ 0.1.}. The relative amount of
events where the $\omega$\ mesons decay at finite nuclear matter
density is vanishingly small. The fraction of the 'inside' decays
increases up to 1/3  for the $\omega$ mesons of momentum $\leq$ 0.3
GeV/c relative to the projectile nucleus frame of reference (middle
panel of Fig.~\ref{Fig_Om_Inv_Mass}). The mass distribution for low
momentum $\omega$'s produced in the central collisions with the
impact parameter $b<R_{Cu}/2$ is shown in the right panel of
Fig.~\ref{Fig_Om_Inv_Mass}. One can see further drop of the vacuum
peak accompanied by the enlargement of the 'inside' component up to
almost 2/3. The position of the vacuum peak can be used as a
reference point on the invariant mass scale.

\begin{figure}
\begin{center}
\includegraphics[width=12cm]{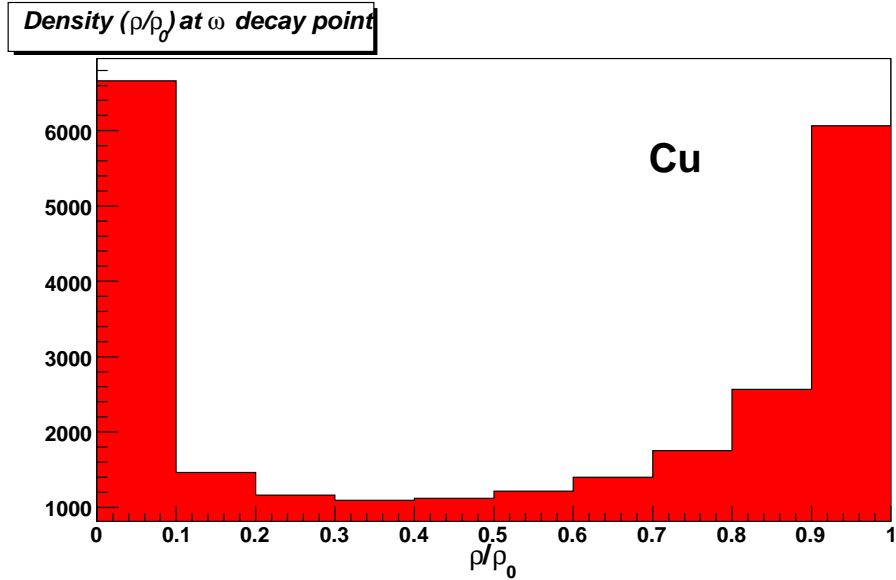}
\caption{Nuclear density at the $\omega$ meson decay point for $P <
0.3 GeV/c$ and $b<R_{Cu}/2$} \label{Fig_Rho_Om_Dec_1}
\end{center}
\end{figure}

The density distribution with the above cuts on the $\omega$
momentum and the impact parameter is shown in
Fig.~\ref{Fig_Rho_Om_Dec_1}. It is seen that significant part of low
momentum mesons produced in the central collisions decays in dense
layers of the nucleus. Note that the calculations within the folding
model provide the possibility to estimate average nuclear density
for the 'inside' decay component.
It should be mentioned that inelastic $\omega$-nucleus collisions,
i.e., the processes $\omega N\rightarrow \omega X$ result in slowing
down of the $\omega$ mesons and an enhancement of low momentum part
of the spectrum. Moreover, one can think that the range of low
$\omega$ momenta would be even further enhanced due to decrease of
the absorption effect. The value of $\sigma_{\omega N}$ in nuclear
matter is expected to be less than one in the free space since Pauli
blocking prevents the low energy $\omega$-nucleon interactions.

Thus, we conclude that the prospective signal of the $\omega$ meson
mass shift is strong enough to be observed experimentally. The
proposed detector layout will permit us to collect a large amount of
the $\omega$ mesons with low momenta relative to the surrounding
nuclear matter (see section~\ref{EventRate}) and study the momentum
dependence of the predicted effect.

\subsection{In-medium $\omega$ meson width}
\label{InMediumWidth}

The straightforward determination of the in-medium $\omega$ meson
width from the shape of the observed mass spectrum is hardly
possible because the rescattering of the pion would changes its
kinematical parameters which results in distortion of the observed
invariant mass peak. The authors of Ref.~\cite{Cabrera} have
proposed the alternative method to study the $\phi$-meson width in
the nuclear medium - by an attenuation measurements of the $\phi$
meson flux in photonuclear reactions on different nuclear targets.
This method is based on the well known connection between the
particle absorption in nucleus and the imaginary part of the
respective nuclear optical potential (see Eq.\ref{eq:Param}). The
method proposed in Ref.~\cite{Cabrera} was applied by Muehlich and
Mosel to the $\omega$ photo-production ~\cite{MuehlichMosel}. The
flux of $\pi^{0}\gamma$ pairs which escape a nucleus had been
calculated within the Boltzmann-Uehling-Uhlenback (BUU)
coupled-channel transport approach. As a measure of the $\omega$
width in nuclei the authors of Ref.~\cite{MuehlichMosel} used the
so-called nuclear transparency ratio:
\begin{equation}
\label{eq:TransparencyRatio}
   T_{A} = \sigma_{\gamma A
\rightarrow VX}/A \sigma_{\gamma N \rightarrow VX},
\end{equation}
i.e. the ratio of the inclusive $\omega$ photo-production cross
section on nucleus divided by A times the same quantity on a free
nucleon. It can be interpreted as the probability of the $\omega$
meson to get out of the nucleus. It was shown that the A-dependence
of the production cross sections significantly differs from that
expected in the case when there are no medium effects on the
$\omega$ width. Similarly, the valuable information about the
$\omega$ width in the matter can be obtained from the analysis of
A-dependence of $\omega$ meson production cross section in
proton-induced reactions.

Although a proton initial state interaction is rather strong, the
$\omega$ absorption is essential. For small angle $\omega$
production the last effect can be taken into account by the Glauber
eikonal factor which explicitly depends on the $\omega$ meson width
$\Gamma^{*}$~\cite{EikonalFactor}:

\begin{equation}
\label{eq:SurvivalProbab}
         P = exp[-\int\limits_0^\infty dl \Gamma^{*}(p_{\omega},\rho(r'))/\beta_{\omega}],
\end{equation}
where $\vec r^{'}=\vec r+l\vec p_{\omega}/|\vec p_{\omega}|$ with
$\vec r^{'}$ the $\omega$ production point, $p_{\omega}$ and
$\beta_{\omega}$ are the momentum and velocity of the $\omega$ in
the target nucleus frame, while $\rho(r')$ stands for the local
nuclear density. Eq.~\ref{eq:SurvivalProbab} shows that the survival
probability P of the $\omega$ meson in its way out of a nucleus
decreases with increasing of the $\omega$ width $\Gamma^{*}$.

The $\omega$ width in nuclear matter is defined by
Eq.\ref{eq:SpectrFunction}, where $\Gamma_{0}$ is free meson width
and the additional width $\Gamma^{*}$ can be expressed according to
Ref.~\cite{MuhMos} as:

\begin{equation}
\label{eq:MuMos}
    \Gamma^{*} = \gamma \{\beta \sigma^{*}_{\omega N}\}\rho(r).
\end{equation}

Here $\beta$ is the the relative velocity of nucleon and $\omega$
meson, $\gamma$ denotes the Lorentz factor for the transformation
from nuclear rest frame to the $\omega$ rest frame, $\rho(r)$ stands
for the local nuclear density. It is seen that in-medium $\omega$
meson width depends on its velocity (momentum) relative to the
nucleus rest frame. To evaluate the sensitivity of the A-dependence
to the magnitude of in-medium $\sigma^{*}_{\omega N}$ we use the
total $\omega N$ cross section in the free space adopted from the
model~\cite{Lykasov}:

\begin{equation}
\label{eq:Lykasov}
    \sigma_{el}=[5.4+10exp(-0.6|\vec q|)]~mb ,
\end{equation}
\begin{equation}
\label{eq:Lykasov1}
 \sigma_{in}=[20+4/|\vec q|]~mb ,
\end{equation}
where q is $\omega$ meson momentum. The brackets in
Eq.\ref{eq:MuMos} indicate an average over the Fermi motion of the
nucleons. In Ref.~\cite{MuehlichMosel} the $\omega$ collision width
was estimated as 37 MeV at nuclear saturation density for vanishing
meson momentum.

\begin{figure}[h]
\begin{center}
\includegraphics[width=12cm]{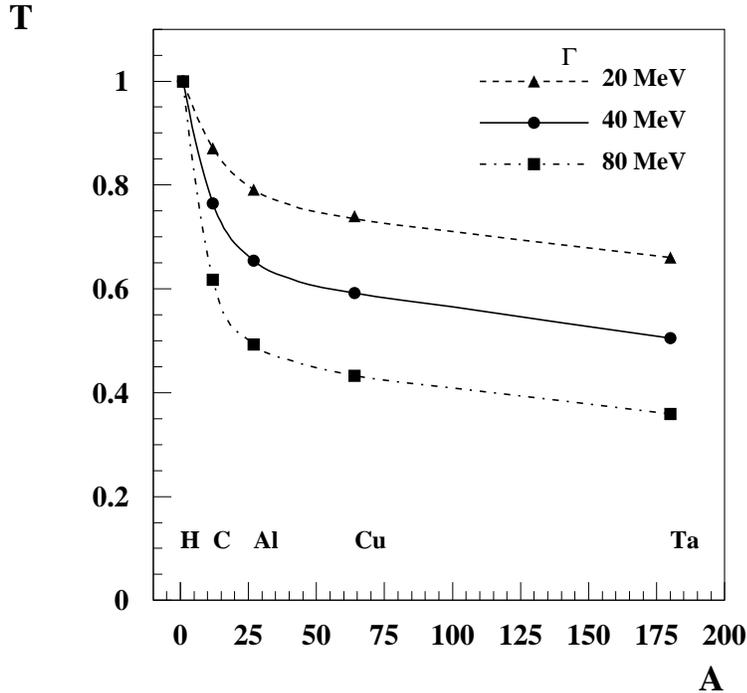}
\caption{Transparency as a function of atomic mass number}
\label{Fig_transparency}
\end{center}
\end{figure}

\begin{figure}[h]
\begin{center}
\includegraphics[width=12cm]{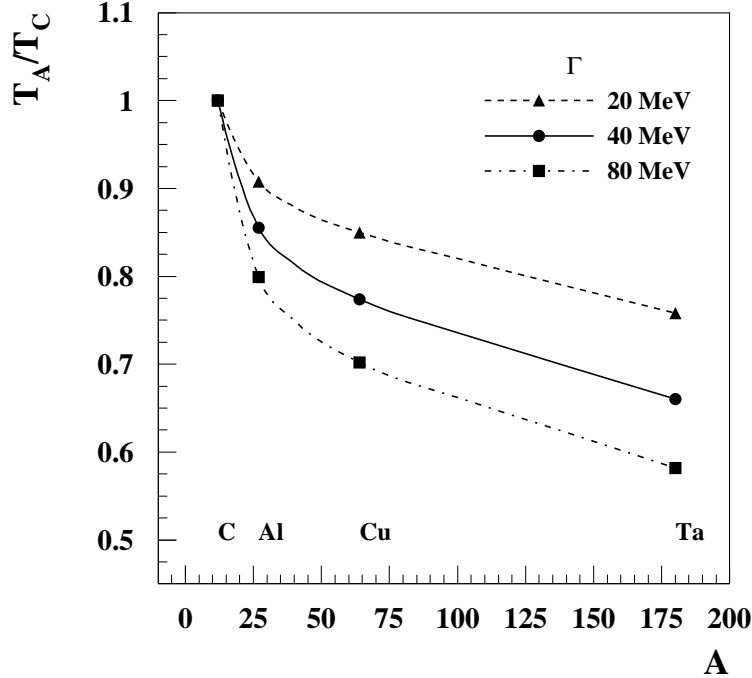}
\caption{Transparency ratio as a function of target mass number}
\label{Fig_transp_ratio}
\end{center}
\end{figure}

The momentum averaged atomic mass dependence of the transparency
obtained within the folding model is shown by solid curve in
Fig.~\ref{Fig_transparency}.
The dash and dash-dotted lines - which correspond to the
calculations with the value of $\Gamma^{*}$ multiplied by a factors
of 0.5 and 2, respectively, - reflect the sensitivity of the
A-dependence to the collision width $\Gamma^{*}$. The difference
between the curves is large enough to be detected in the experiment.
Since the shape of the A-dependence, and not so much the absolute
value, is important to learn about the $\omega$ width, the cross
section for middle and heavy nuclei can be normalized to the cross
section for light nucleus. The A-dependence of the transparency
normalized to the cross section for the $\omega$ production on
carbon target is presented in Fig.~\ref{Fig_transp_ratio}.
In spite of some less difference between the curves the measurement
of the ratios has an advantage to cancel out most of systematic
uncertainties. Thus, the calculations clearly show that the proton
induced $\omega$ meson production in nuclei can indeed be used to
get information on the $\omega$ width in the medium.

Obviously, the discussed above $\omega$ width calculated according
the Eq.~\ref{eq:MuMos} with total $\omega$-nucleon cross section in
the free space is only the estimate. The real total $\omega$N cross
section in the medium can differ from that in the free space.
Indeed, the experiment on incoherent $\phi$ photoproduction on Be,
C, Al and Cu targets recently performed at SPring8
(LEPS)~\cite{Ishikava} has found an unexpectedly strong dependence
of the loss of $K^{+}K^{-}$ flux from the $\phi$ decay on target
mass number. The total in-medium $\phi N$ cross section has been
estimated by the authors as $35^{+17}_{-11}$ mb using the
Glauber-type multiple scattering theory. This value significantly
differs from $\sigma^{tot}_{\phi N}$ in the free space which is
equal to 9-11 mb. One can expect similar effect for the $\omega$
mesons.

Note that the investigation of the $\omega$ meson width and its
momentum dependence in pA reactions has several important advantages
as compared with one in Ap reactions. First, the proton beams are
usually much more intensive than the ion ones. Second, the simple
solid targets can be used instead of the hydrogen target. Third, in
the pA collisions the momentum dependence of the $\omega$ meson
width can be studied in wide momentum range from 0.5 to 4 GeV/c. It
is worth to note, that the possible effect of density dependent
$\omega$ meson mass shift is of minor importance in high momentum
range.

\section{Simulations with RQMD event generator}
\label{simulations}

For the simulations we use Relativistic Quantum Molecular Dynamics
(RQMD)~\cite{RQMD} event generator, version 4.12. RQMD produces
hadrons through the excitation of baryonic and mesonic resonances.
Heavy resonances (more than 2 GeV for baryons and more than 1 GeV
for mesons) are treated in the string picture following the Lund
model \cite{LUND} and all particles are allowed to reinteract
(baryon-baryon, baryon-meson and meson-meson). The model provides a
complete time-dependent description of the evolution of each event.
The probabilities for excitation of specific channels are governed
by experimental cross sections to the extent possible. The formation
points of hadrons are taken from the properties of resonance decay
and string fragmentation.

About $3 \times 10^{6}$ minimum bias Cu+p events have been
generated. The following simulation was performed for the ring-like
electromagnetic calorimeter covering the range of the polar angles
$\theta = 5^{0}-25^{0}$ and full azimuthal angle of $0^{0} < \phi <
360^{0}$. At projectile energy of 4 AGeV the mean total multiplicity
is equal to 12 for Cu+p and 3.4 for C+p collisions. Multiplicities
of different species are presented in Table~\ref{tab_Mult}. Since
the charged component (protons and pions) amounts to one half of the
total multiplicity one can estimate the impact parameter of the
collision by detecting the charged ejectiles by CPV counters.
\begin{table}[tbp]
\begin{center}
\begin{tabular}{|c|c|c|c|c|c|c|c|}
\hline
species     & n & p &$\pi^{o}$&$\pi^{+}$&$\pi^{-}$&$\eta$&$\gamma$ \\
\hline
multiplicity&4.7&4.6&   1.47  &   0.74  &   0.66  & 0.037& 0.0026  \\
\hline
\end{tabular}
\end{center}
\caption{Mean multiplicities of particles predicted by the RQMD code for the
minimum bias Cu+p events at 4 AGeV}
\label{tab_Mult}
\end{table}
\begin{figure}[h]
\begin{center}
\includegraphics[width=12cm]{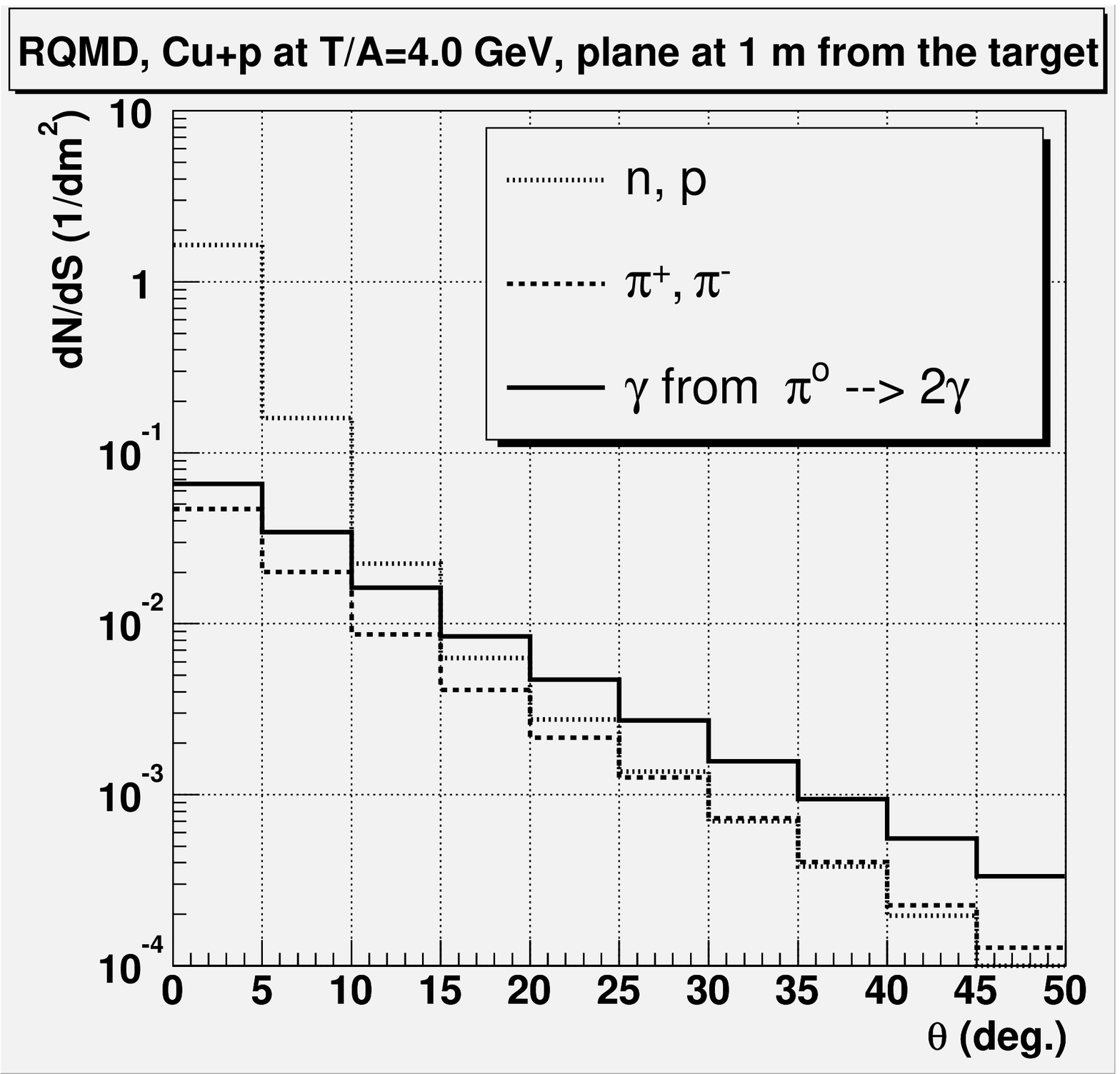}
\caption{Density
of particles in a plane at 1 m from the target}
\label{Fig_occupancy}
\end{center}
\end{figure}

The angular dependence of the secondaries in the laboratory is shown
in Fig.~\ref{Fig_occupancy}. It is seen that the multiplicity drops
rapidly with the production angle. Due to the ring-like geometry of
the EMCAL the first angular bin $0^{0}-5^{0}$ is out of the detector
acceptance. Assuming the target-detector distance of 1 meter and
granularity of the EMCAL of $2 \times 2  cm^{2}$ one gets for the
interval of $5^{0}-10^{0}$ maximum cell occupancy of about 0.006 for
minimum bias events and approximately two times more for most of
central collisions.

\begin{figure}[p]
\begin{center}
\includegraphics[width=12cm]{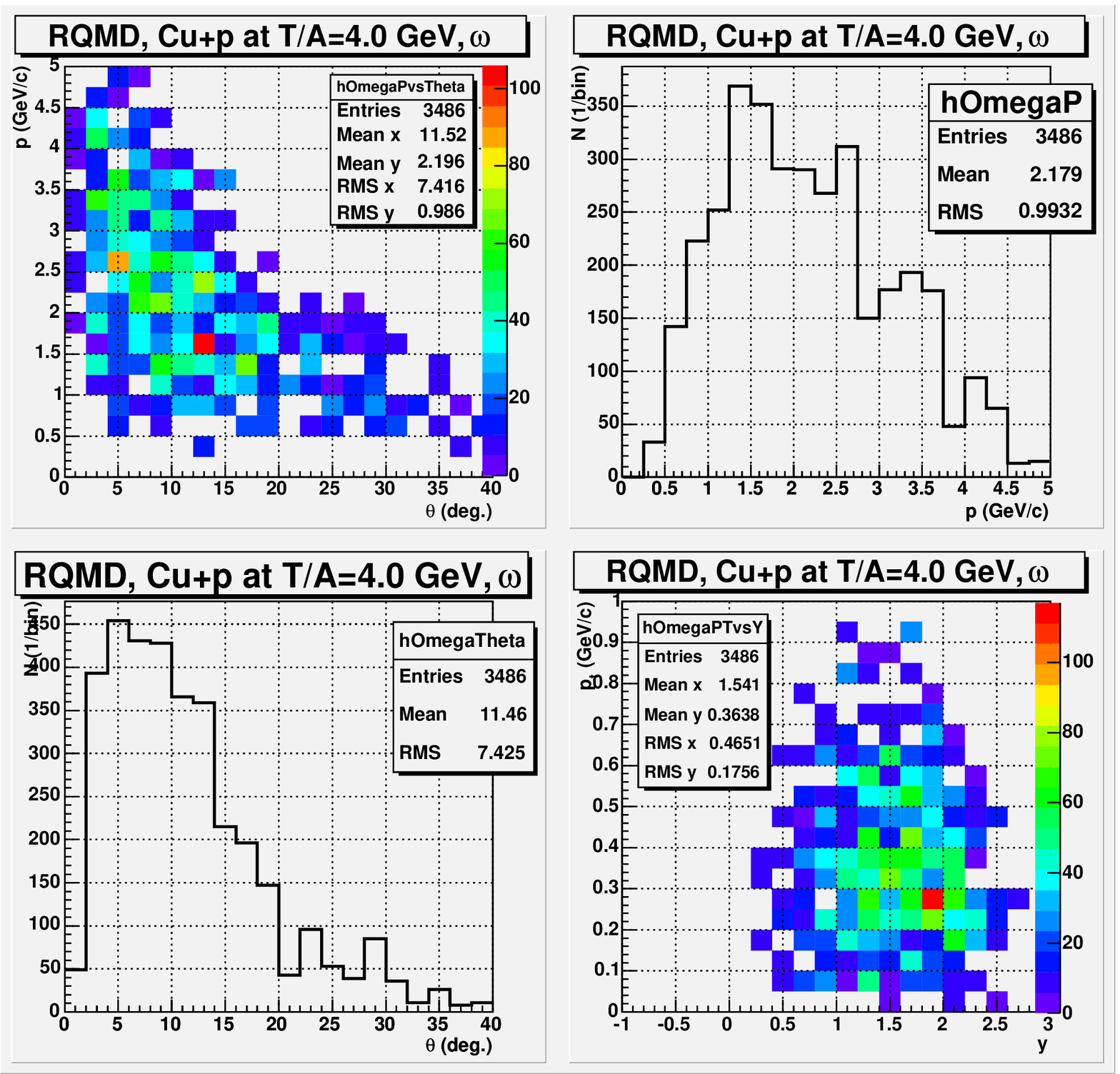}
\caption{$\omega$ spectrum in the laboratory frame}
\label{Fig_pOmegaAnal}
\end{center}
\end{figure}

\begin{figure}[h]
\begin{center}
\includegraphics[width=12cm]{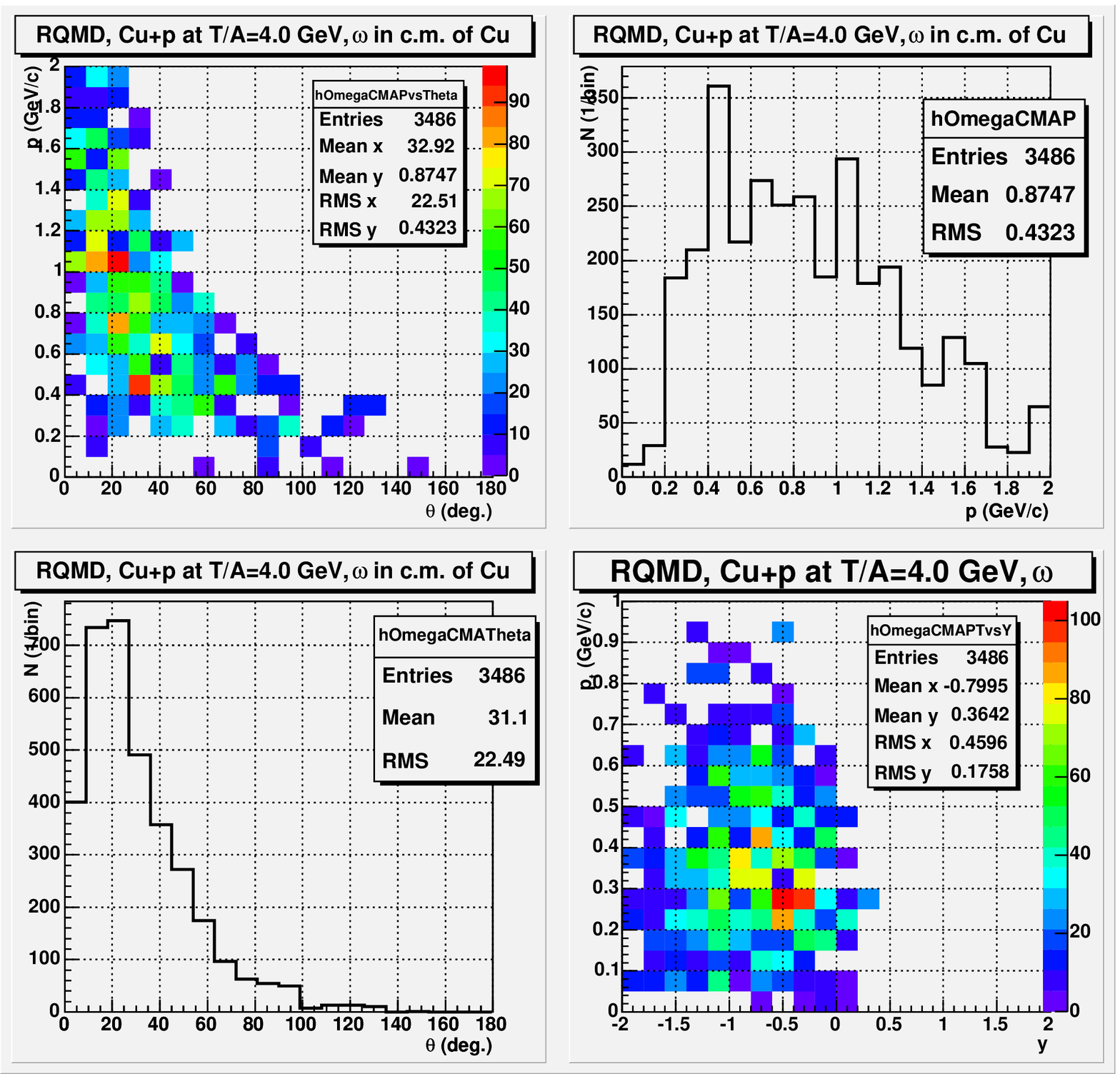}
\caption{$\omega$ spectrum in the projectile nucleus frame}
\label{Fig_pOmegaAnalCMA}
\end{center}
\end{figure}

The two-dimensional plot for the $\omega$ mesons produced in Cu+p
collisions at ion beam energy of 4 AGeV is presented in the left
upper panel of Fig.~\ref{Fig_pOmegaAnal}.
\begin{figure}[p]
\begin{center}
\includegraphics[width=15cm]{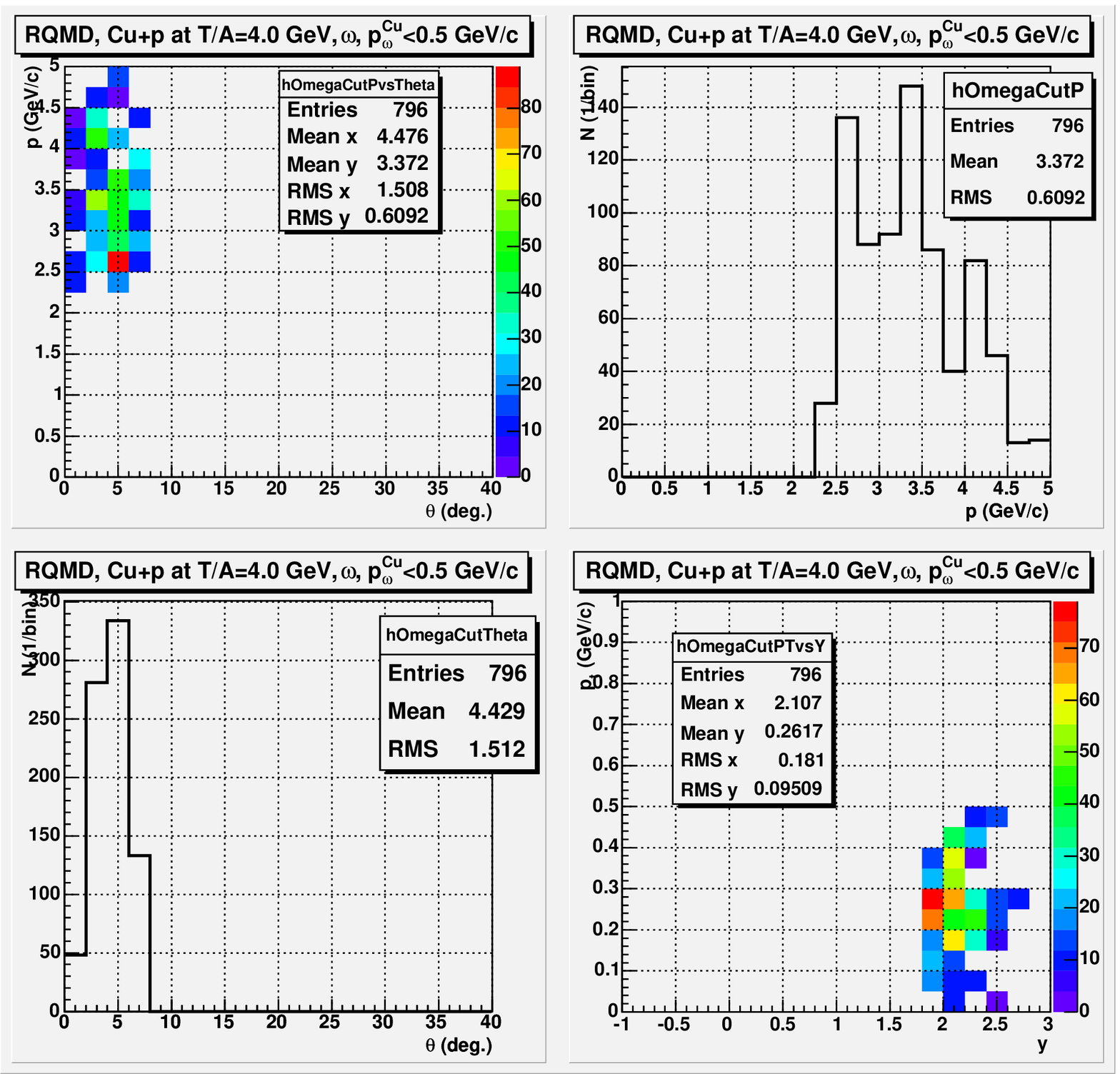}
\caption{The same as for Fig.~\ref{Fig_pOmegaAnal}
but for $\omega$ with $p<0.5$ GeV/c in the projectile nucleus frame}
\label{Fig_pOmegaCutAnal}
\end{center}
\end{figure}
It is seen that fast mesons in the laboratory (the target proton
frame of reference) are predominantly concentrated in the range of
small angles. The momentum spectrum of all produced $\omega$'s
(right upper panel of Fig.~\ref{Fig_pOmegaAnal}) extends up to 5
GeV/c. In the right upper panel of Fig.~\ref{Fig_pOmegaCutAnal} we present
the same laboratory spectrum with cut on the $\omega$ meson momentum
$p<$ 0.5 GeV/c in the incident nucleus rest frame. The comparision
of the right upper panels of Fig.~\ref{Fig_pOmegaAnal} and
Fig.~\ref{Fig_pOmegaCutAnal} shows that almost 1/4 part of all
produced mesons are within the range $p<$ 0.5 GeV/c in the
projectile Cu frame of reference. In the right bottom panels of
Fig.~\ref{Fig_pOmegaAnal} and Fig.~\ref{Fig_pOmegaCutAnal} the
two-dimensional distributions - rapidity versus transverse momentum
- are presented without and with the above cut on the $\omega$
momentum, respectively. One can see the sizeable number of events
with small $p_{t}$ in the vicinity of the projectile nucleus
rapidity $Y$=2.34 in the laboratory system. The most of $\omega$'s
have the rapidities $Y<$ 2.34 and show up at $Y<$ 0 in the
projectile nucleus rest frame (bottom right panel of
Fig.~\ref{Fig_pOmegaAnalCMA}). Thus, the range of low $p_{t}$ and
very small rapidities is accessible for the investigation in Ap
kinematics. As was above mentioned  the $\omega$ meson mass shift
probably depends on the $\omega$ momentum relative to the nuclear
medium. The strongest effect is expected for the momenta of less
than 0.3 GeV/c. The study of the $\omega$ meson invariant mass
distributions reconstructed for different momentum bins in the range
of $P_{\omega} <$ 0.5 GeV/c provides the possibility to explore the
momentum dependence of the $\omega$ mesons mass shift in the medium.

For the selection of useful events the EMCAL should be able to
detect efficiently three photons from the $\omega \rightarrow
\pi^{0}\gamma \rightarrow 3\gamma$ decay. The performed simulations
show that the geometrical efficiency for the detection of the
photons from the $\omega$ decay is near 40 \%
for the range of small production angles ($\theta<7.5^{0}$) and high
$\omega$ momentum ($p>2.25 GeV/c$) in the laboratory which
corresponds to the range of $p_{\omega}<$ 0.5 GeV/c in the
projectile nucleus rest frame. Since invariant mass resolution
decreases with lowering of the photon energy,
we have determined the efficiency after applying the cut on
$E_{\gamma}>$ 0.5 GeV. It amounts to about 15-20\% in the range of
our interest. The simulations indicate that in Ap kinematics the
detection efficiency of low momentum $\omega$ mesons relative to the
projectile nucleus turns out approximately by an order of magnitude
higher than that averaged over all meson momenta.

As was explained in section~\ref{InMediumWidth} the in-medium
$\omega$ meson width can be deduced from the analysis of the
A-dependence of the production cross sections of relatively fast
$\omega$'s, which are mostly decay outside the nucleus. Obviously,
that the photons from the decay of high momentum mesons produced in
traditional pA kinematics will also be detected with high efficiency
because the efficiency depends on photon energy relative to the
detector.

Thus, the momentum dependencies of both in-medium $\omega$ meson
mass and width can be investigated in the inverse and direct
kinematics -- i.e. using the ion and proton beams -- without the
change of the detector position and its layout.

\section{Background and its suppression}
\label{background}

The feasibility of the experiment depends on the signal to
background ratio. RQMD simulations show that the main source of the
background is the $\pi^{0}\pi^{0}$ production. Such events can lead
to the misidentification due to a finite geometry of the detector if
one of the four photons is out of the EMCAL acceptance. The
contribution from other sources of the background like
$\eta\pi^{0}$, $\eta^{'}$, $\Delta^{0} \rightarrow n\gamma$ etc. is
relatively small in the invariant mass range of interest 0.65 - 0.85
GeV since the invariant masses reconstructed from kinematical
parameters of three uncorrelated photons are spread over wide mass
range from 0.1 to 1.0 GeV. Therefore, the useful events from the
$\omega \rightarrow \pi^{0}\gamma \rightarrow 3\gamma$ decay will be
detected on the top of smooth continuum steming mainly from the
$\pi^{0}\pi^{0}$ production process. The Signal/(Signal+Background)
ratio $R=S/(S+B)$ for the minimum bias events is less than
approximately one per cent. However, this value can be significantly
improved by applying the appropriate kinematical cuts. RQMD
simulations indicate that the spectrum of the photons originating
from the $\pi^{0}$ decay drops  steeper than that from the
$\omega$ decay. By this reason the cut on photon energy should lead
to the background suppression.
\begin{figure}[p]
\begin{center}
\includegraphics[width=16cm]{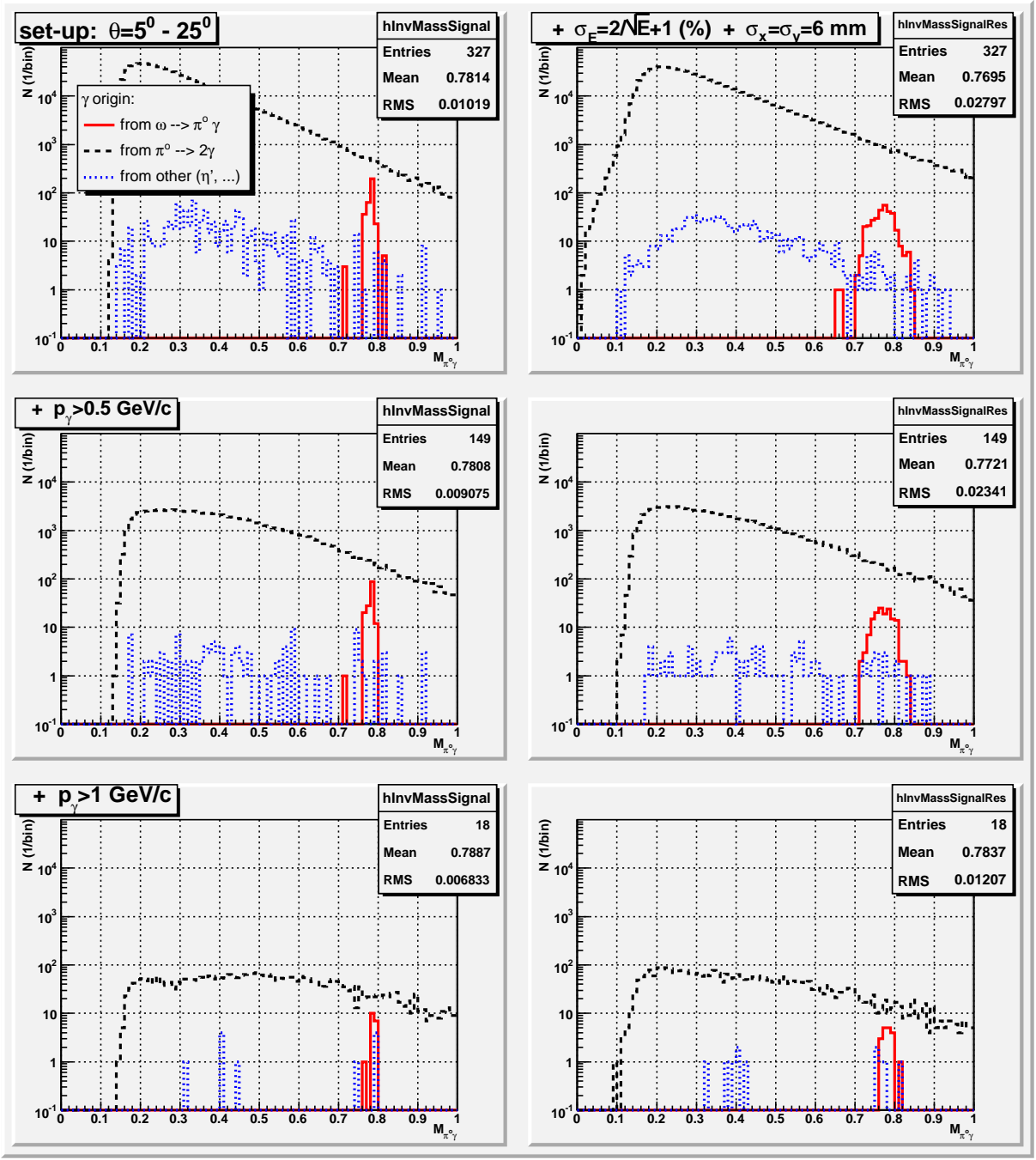}
\caption{Invariant mass spectra with and without cuts on the photon
energies for RQMD Cu+p events at 4 AGeV}
\label{Fig_cInvMass}
\end{center}
\end{figure}
The histograms in Fig.~\ref{Fig_cInvMass} demonstrate this
effect\footnote{The medium effects are not included in the RQMD
simulations}. Invariant mass distributions of $\pi^{0}\gamma$ system
without the cut on photon energies are shown in the upper row, while
the middle and bottom rows represent the distributions obtained with
cuts on $E_{\gamma}>$ 0.5 GeV and $E_{\gamma}>$ 1 GeV,
correspondingly. The mass distributions from RQMD simulations are
presented in the left column. In the right column the same
distributions are depicted for the case when the energy and space
resolution of the PbWO calorimeters are taken into account. The
numbers of the signal (S) and signal plus background (S+B) events
are collected in Table~\ref{tab_InvMass}. The magnitude of R=21\%
and mass resolution of 12 MeV can be reached after implying the cut
on a photon energy $E_{\gamma}>$ 1 GeV. However, such cut leads to
the reduction of signal events by a factor of 8 compared to the case
$E_{\gamma}>$ 0.5 GeV.
\begin{table}[tbp]
\begin{center}
\begin{tabular}{|c|c|c|c|c|}
\hline
  EMCAL&  info    &         all         &$E_{\gamma}>0.5 GeV$ &$E_{\gamma}>1 GeV$\\
\hline
without&S/(S+B)            &9.5\%       &10.5\%                  &21\% \\
resol. &RMS(S) MeV&        10           &         9           &       7        \\
\hline
PbWO   &S/(S+B)            &1.8\%       & 5.9\%                 &21\% \\
resol. &RMS(S) MeV&        28           &        23           &      12        \\
\hline
\end{tabular}
\end{center}
\caption{Parameters of signal (S) and background (B) $\pi^{o}\gamma$
pairs for RQMD Cu+p events at 4 AGeV} \label{tab_InvMass}
\end{table}
The discussed Signal/(Signal+Background) ratio can be even further
improved since the $\pi^{0}$ and $\gamma$ which are, in fact, from
the $\omega$ decay are strongly correlated while the photons
steming from the $\pi^{0}\pi^{0}$ or other sources will not show
such a correlation. Due to two-body nature of
the $\omega \rightarrow \pi^{0}\gamma$ decay the pion should be
emitted in the plane which is formed by the projectile momentum and
the momentum of the photon originated from the $\omega$ decay. In
such a case the difference in the azimuthal angles of the $\pi^{0}$
and $\gamma$ should be close to 180 degrees. The cut
$\phi_{\pi}-\phi_{\gamma}>100^{0}$
results in additional increase of R by a factor of 2. Thus, we
conclude that the applying of the appropriate cuts will permit to
reduce the background to the level acceptable for the measurements.

\section{Event trigger}
The events of interest detected by EMCAL are accompanied by the
deposit of energy in three or more groups of cells. For the
selection of such events a multilevel trigger could be used. A first
level should be provided by the signals selected the events with
energy deposit of more than 0.5 GeV for the suppression of
background from low momentum $\pi^{0}$ production and their
rescattering. The second level of the trigger would include the
requirement the energy deposit of more than 1 GeV in one of the cell
group which correspond to the photon energy from the real $\omega$
meson decay. The third level would select the events with large
azimuthal separation corresponding to actual two-body $\omega$
decays. One can estimate a contribution of background events using
the experimental counting statistics of the inclusive trigger and
the trigger of delayed coincidence. The information from CPV
counters will be used as an off-line trigger for the reduction of
charged background and for the estimation of the collision
centrality.

\section{Event rate estimate}
\label{EventRate}

Let us first estimate the expected number of events for the
production of low momentum mesons with respect to a projectile
nucleus. Assuming a moderate ion beam intensity of $1 \times 10^{8}$
ions/cycle, the extraction efficiency of 50\% and target efficiency
of 2\% one gets the number of the ion interactions inside the target
of $1 \times 10^{6}$ per one accelerator cycle. The normalization
factor $N$ - which is the ratio of the number of interactions during
one accelerator cycle to the number of simulated collisions  - is
equal to 0.33. The momentum intervals of less than 0.3 GeV/c and 0.5
GeV/c contain 225 and 796 events, respectively (see right top
histogram in Fig.~\ref{Fig_pOmegaAnalCMA}).

Assuming the cycle repetition of 10 $min^{-1}$ and taking into
account the detector efficiency of 15\% (after applying the cut on
$E_{\gamma}>$ 0.5 GeV) one can estimate the event rate for the
above momentum ranges as:

     N($P_{\omega} <$ 0.3 GeV/c) = 0.33 x 225 x 0.15 x 10 = 111 events/min

     N($P_{\omega} <$ 0.5 GeV/c) = 0.33 x 796 x 0.15 x 10 = 398 events/min.

As was described in the section~\ref{background} one has to apply
several cuts to reach the background conditions acceptable for the
measurements. That leads to the reduction of useful events by a
factor of 10-20. In the most pessimistic case the numbers of useful
events collected during one day measurement are:

        N($P_{\omega} <$ 0.3 GeV/c) =  $8 \times 10^{3}$ events/day

        N($P_{\omega} <$ 0.5 GeV/c) =  $28.6 \times 10^{3}$  events/day

The number of events in the lowest momentum interval $P_{\omega} <$
0.1 GeV/c is equal to approximately 350 per day. For the carbon
projectile the event rates will be less by a factor of 4-5 due to
the A-dependence of the production cross section. Thus, the estimate
clearly demonstrates that the investigation of the momentum
dependence of the expected $\omega$ meson mass shift in the nuclear
matter can be performed with high statistical accuracy.

It should be noted that the above estimates are based on the RQMD
simulations which disregard any medium effects on the $\omega$. If
the $\omega$ meson mass really drops in nuclear matter by 80-100 MeV
the production cross section would increase due to the downward
shift of the reaction threshold. Moreover, one can expect that the
produced mesons will be decelerated during their way out of the
nucleus due to the action of the attractive nuclear $\omega$ meson
potential. This slowing down would lead to the increase in number of
events in low momentum range. Note also that the energy loss of a
meson in the elastic and quasielastic $\omega N$ scattering inside a
nucleus results in the same effect.

The usage of more intensive proton beam provides the possibility to
collect large amount of data on high momentum $\omega$ meson
production and perform the detail investigation of the momentum
dependence of the $\omega$ meson width in the nuclear matter. Data
on the $\omega$ meson production in the momentum range around 1.5
GeV/c can be obtained in both inverse and direct kinematics. These
data will be used for the cross check and mutual normalization of
the Ap and pA measurements. The statistics which can be obtained in
AA interactions is obviously higher than that in Ap collisions.

\section{Conclusion}

The modification of the properties of the vector mesons in baryon
environment continue to be one of the most interesting topics in
hadron physics today.

We suggest to investigate the in-medium properties of the $\omega$
mesons at normal nuclear density in nucleus-proton and
proton-nucleus collisions as well as at higher density in
nucleus-nucleus collisions at ITEP accelerator facility TWAC.

Study of the Ap and pA reactions is the effective tool to get the
information on the in-medium $\omega$ meson properties at normal
nuclear density. The using of the inverse Ap kinematics and
$\omega\rightarrow\pi^{0}\gamma$ decay mode permits to collect large
statistics for production of the $\omega$ mesons with low momenta
relative to the nuclear matter. Estimated high event rate offers the
possibility to split the statistics into several momentum bins and
study the $\omega$ meson mass shift in wide momentum interval {\sl
including not yet explored range of momentum less than 0.3 GeV/c
which is expected to be most sensitive to the mass change effect}.
The detail information on in-medium $\omega$ meson width in wide
momentum interval will be obtained in pA collisions. The goal of the
first stage of the experiment is to make decisive conclusion about
the in-medium $\omega$ meson mass and width at normal nuclear
density.

On the second stage of the suggested study we shall obtain the
information on the $\omega$ meson production in nucleus-nucleus
collisions which will be used for the investigation of the in-medium
$\omega$ meson properties at higher density compared to that
accessible in nucleus-proton interactions. The results obtained at
the first stage of the investigation at normal nuclear density will
provide the reliable basis for the selection and interpretation of
specific nucleus-nucleus phenomena.

Ap, pA and AA measurements will be performed in quite identical
conditions using the same experimental set-up.

\section{Further investigations at ITEP and GSI}

One of the possible extension of the proposed studies in a few GeV
energy range is the investigation of the in-medium change of
unflavored mesons. The performed RQMD simulations of the Ap
collisions show the significant yield of the $\eta$ and $\eta{'}$
mesons for which the modification effects had also been
theoretically predicted ~\cite{Hagahiro},~\cite{Review}. The
branching ratios of the $\eta\rightarrow\gamma\gamma$ and
$\eta{'}\rightarrow\pi_{0}\pi_{0}\gamma\gamma$ decays are as high as
39\% and 21\%, correspondingly. The properties of these mesons can
be investigated at TWAC using the same experimental set-up.

The in-medium properties of the charmed mesons and charmonium can be
explored at significantly higher ion energy which will be accessible
at the new FAIR facility (GSI). In particular, the
 mass splitting of $\overline{D}D$ mesons at high baryonic density ~\cite{DDsplitting}
 will be investigated by CBM experiment in heavy-ion collisions.
 As a masses of charmonia are large, only little sensitivity to
changes in the quark condensate is expected. Consequently, the
in-medium mass of the charmonium states would be affected primary by
a modification of the gluon condensate. Large attractive mass shifts
are predicted for exited charmonium states ~\cite{CharmShift}. Using
the inverse and direct kinematics provides the possibility to study
the production and propagation of both low and high momentum heavy
quark systems in baryonic matter.

\section{Acknowledgments}
The authors gratefully acknowledge N.O.Agasian, B.L.Ioffe,
L.A.Kondratuyk for fruitful discussions. This work was partially
supported by Federal agency of Russia for atomic energy (Rosatom).

--- {{\bf The suggested investigation is open for the cooperation with the
experimentalists and theorists who are interested in the above
discussed physics and in the performing of the proposed
experiment.}}

\end{document}